\documentclass[aps,prd,floatfix,groupedaddress,nofootinbib,superscriptaddress,twocolumn,notitlepage]{revtex4-1}

\usepackage{amsmath,amssymb,bm,color,float,graphicx,mathrsfs}
\usepackage[hidelinks]{hyperref}
\hypersetup{colorlinks=true,linkcolor=blue,citecolor=blue,urlcolor=blue}
\usepackage{Macro}

\def\hCXX{\widehat{C}^{\rm XX}}

\begin{document}

% Title %
\title{Tomographic constraint on anisotropic cosmic birefringence}

\author{Toshiya Namikawa}
\affiliation{Department of Applied Mathematics and Theoretical Physics, University of Cambridge, Wilberforce Road, Cambridge CB3 0WA, United Kingdom}
\affiliation{Center for Data-Driven Discovery, Kavli IPMU (WPI), UTIAS, The University of Tokyo, Kashiwa, 277-8583, Japan}

% Date %
\date{\today}

% Abstract %
\begin{abstract}
We constrain anisotropic cosmic birefringence generated at reionization using Planck PR4 polarization data for the first time. 
Several recent analyses of WMAP and Planck polarization data have found a tantalizing hint of isotropic cosmic birefringence. Ongoing and future cosmic microwave background (CMB) experiments will test isotropic cosmic birefringence by improving the absolute angle calibration and understanding the intrinsic parity-odd power spectrum of the Galactic foregrounds. Alternatively, measuring anisotropies in cosmic birefringence and its time evolution is also a key observable to confirm the signal of cosmic birefringence and to investigate its origin. We discuss estimators of anisotropic cosmic birefringence generated at different redshifts. We then estimate anisotropic cosmic birefringence generated at reionization from the PR4 data, showing that the power spectrum is consistent with null. We find that the model proposed by Ferreira et al. is still consistent with the observation. Future full-sky CMB experiments such as LiteBIRD and PICO will help tighten the tomographic constraint to test models of cosmic birefringence. 
\end{abstract} 


%////////////////////////////////////////%
% MAIN MATTER 
%////////////////////////////////////////%

% Contents %
\maketitle

%////////////////////////////////////////%
\section{Introduction} \label{sec:intro}
%////////////////////////////////////////%

Recent analyses of cosmic microwave background (CMB) data have found a tantalizing hint of {\it cosmic birefringence}---a phenomenon in which the polarization plane of light rotates as it travels through space \cite{Minami:2020:biref,Diego-Palazuelos:2022,Eskilt:2022:biref-freq,Eskilt:2022:biref-const,Eskilt:2023:EDE} (see Ref.~\cite{Komatsu:2022:review} for a review). As a parity-violating signature, cosmic birefringence serves as a potential smoking gun for new physics beyond the $\Lambda$ Cold Dark Matter ($\Lambda$CDM) model and the Standard Model of particle physics~\cite{Nakai:2023}. 

Cosmic birefringence can arise from a pseudoscalar field, such as axionlike particles (ALPs), coupled to the electromagnetic field via the Chern-Simons term:
\al{
    \mathcal{L}\supset -\frac{1}{4}g_{\phi\gamma}\phi F^{\mu\nu}\tilde{F}_{\mu\nu}
    \,,
} 
where $g_{\phi\gamma}$ is the coupling constant, $\phi$ is an ALP field, $F^{\mu\nu}$ denotes the electromagnetic field tensor, and $\tilde{F}_{\mu\nu}$ is its dual. Multiple studies have explored cosmic birefringence induced by the ALP field of dark energy \cite{Carroll:1998:DE,Liu:2006:biref-time-evolve,Panda:2010,Fujita:2020aqt,Fujita:2020ecn,Choi:2021aze,Obata:2021,Gasparotto:2022uqo,Galaverni:2023}, early dark energy \cite{Fujita:2020ecn,Murai:2022:EDE,Eskilt:2023:EDE,Kochappan:2024:biref}, dark matter \cite{Finelli:2009,Sigl:2018:biref-sup,Liu:2016dcg,Fedderke:2019:biref,Zhang:2024dmi}, and topological defects \cite{Takahashi:2020tqv,Kitajima:2022jzz,Jain:2022jrp,Gonzalez:2022mcx}, as well as by potential signatures of quantum gravity \cite{Myers:2003fd,Balaji:2003sw,Arvanitaki:2009fg}. Looking ahead, future CMB experiments such as BICEP \cite{Cornelison:2022:BICEP3,BICEPArray}, Simons Array \cite{SimonsArray}, Simons Observatory \cite{SimonsObservatory}, CMB-S4 \cite{CMBS4}, and LiteBIRD \cite{LiteBIRD} are expected to significantly reduce polarization noise and enhance sensitivity to cosmic birefringence signals.

% Future directions
To confirm these signals, further work is needed to address the intrinsic parity-odd correlations in Galactic foregrounds, which may be used to calibrate the absolute polarization angle \cite{Huffenberger:2019:dustEB,Cukierman:2022:dust,Hervias-Caimapo:2024}. Another more direct approach is to reduce uncertainties in the absolute polarization angle by improving calibration techniques \cite{SO:wiregrid,Coppi:2022,Ritacco:2023mgi}. Alternatively, observables unaffected by miscalibrated polarization angle can be considered. For example, recent studies indicate that the time evolution of pseudoscalar fields during recombination and reionization significantly alters the CMB polarization power spectra \cite{Finelli:2009,Lee:2013:const,Gubitosi:2014:biref-time,Sherwin:2021:biref,Nakatsuka:2022,Naokawa:2023,Yin:2023:biref,Naokawa:2024xhn,Murai:2024yul}. Measuring the spectral shape of the $EB$ power spectrum could break the degeneracy between the miscalibrated polarization angle and cosmic birefringence signals. Additional constraints on the late-time evolution of ALPs could be obtained by observing the polarized Sunyaev-Zel'dovich effect \cite{Lee:2022:pSZ-biref,Namikawa:2023:pSZ} and galaxy polarization \cite{Carroll:1997:radio,Yin:2024:galaxy}. These tomographic signals can help reduce degeneracies with the miscalibrated polarization angle.

% Anisotropic cosmic birefringence
Anisotropies in cosmic birefringence are another critical observable that can confirm the isotropic signal. If the ALP field exhibits spatial fluctuations, $\delta\phi$, the resulting polarization rotation will be anisotropic (see, e.g., \cite{Carroll:1998:DE,Lue:1999:biref-EB,Caldwell:2011,Lee:2015,Leon:2017,Yin:2023:biref,Ferreira:2023}). Several studies have placed upper limits on these anisotropies by reconstructing the rotation angle \cite{Gluscevic:2012,PB15:rot,BICEP2:2017lpa,Contreras:2017,Namikawa:2020:biref,SPT:2020:biref,Gruppuso:2020,Bortolami:2022whx,BK-LoS:2023,Zagatti:2024}, while others have constrained them using CMB polarization power spectra \cite{Li:2014,Alighieri:2014yoa,Liu:2016dcg,Zhang:2024dmi}. However, time evolution tends to significantly suppress the $B$-mode power spectrum \cite{Namikawa:2024:BB}. 

The time evolution of the ALP field is also essential for constraining models that predict anisotropic birefringence. For instance, Ref.~\cite{Ferreira:2023} proposed a model that produces both isotropic and anisotropic birefringence, with large anisotropic signals generated during reionization. Notably, current measurements of anisotropic birefringence assume it originates during recombination. In this paper, we focus on constraining anisotropic birefringence generated during reionization by constructing an estimator tailored for this epoch. We then apply it to Planck PR4 data to measure the anisotropic cosmic birefringence at reionization.

This paper is organized as follows. In Sec.~\ref{sec:method}, we explain our reconstruction methodology. In Sec.~\ref{sec:datasim}, we describe our data and simulation and test the estimator of anisotropic cosmic birefringence generated during reionization. Section~\ref{sec:results} shows our results for the reconstructed spectrum and the resulting constraint on the scale-invariant birefringence spectrum. We summarize our results and provide a forecast in Sec.~\ref{sec:summary}.

%////////////////////////////////////////%
\section{Tomographic reconstruction of anisotropic cosmic birefrigence} \label{sec:method}
%////////////////////////////////////////%

For a single source plane, such as the last-scattering surface of CMB, the rotation angle at each line-of-sight direction $\hatn$ can be reconstructed from the off-diagonal mode-mode covariance between the $E$ and $B$ modes \cite{Kamionkowski:2009:derot,Namikawa:2020:biref}. The power spectrum of the anisotropic rotation angle can be obtained by squaring the rotation estimator and subtracting relevant biases \cite{Namikawa:2016:biref-est}.
Here, we extend the formalism to the case in which the CMB photons are generated at multiple epochs. 

\subsection{Anisotropic cosmic birefringence from recombination and reionization}

As a simple case, we first consider the case in which the CMB photons are generated from the recombination and reionization epochs. 
The presence of the cosmic birefringence effect rotates the primordial Stokes parameters as \cite{Greco:2022:cross,Greco:2022:aniso-biref-tomography}
%--------------------------------------------------%
\al{
    Q'(\hatn)\pm\iu U'(\hatn) = \sum_{x={\rm rei,rec}}[Q_x(\hatn)\pm\iu U_x(\hatn)]\E^{\pm 2\iu \alpha_x(\hatn)}
    \,. \label{Eq:qurot}
}
%--------------------------------------------------%
Here, $Q_x$ and $U_x$ represent the CMB polarization generated at reionization ($x={\rm rei}$) and recombination ($x={\rm rec}$). Similarly, $\alpha_x$ denotes the rotation angle at reionization ($x={\rm rei}$) and recombination ($x={\rm rec}$). Consequently, the rotation angle modifies the CMB $E$ and $B$ modes. The $E$ and $B$ modes are obtained by transforming the $Q$ and $U$ maps with the spin-2 spherical harmonics, $Y_{\l m}^{\pm 2}$, as \cite{Kamionkowski:1996:eb,Zaldarriaga:1996xe}: 
%--------------------------------------------------%
\al{
    E_{\l m}\pm \iu B_{\l m} = -\Int{2}{\hatn}{}(Y^{\pm 2}_{\l m})^*[Q(\hatn)\pm\iu U(\hatn)]
    \,. 
}
%--------------------------------------------------%
Thus, the $E$ and $B$ modes in the presence of an anisotropic rotation angle are derived by substituting \eq{Eq:qurot} into the above equation and are given up to linear order in $\alpha$ by 
%--------------------------------------------------%
\al{
    E'_{\l m}\pm\iu B'_{\l m} &= E_{\l m}\pm\iu B_{\l m} + \sum_{x={\rm rei,rec}}\sum_{LM\l'm'}\Wjm{\l}{\l'}{L}{m}{m'}{M} 
    \notag \\
    &\qquad\times W^\pm_{\l L\l'} [E^x_{\l'm'}\pm\iu B^x_{\l'm'}] (\alpha^x_{LM})^* 
    \,, 
}
%--------------------------------------------------%
with
%--------------------------------------------------%
\al{
    W^\pm_{\l_1\l_2\l_3} 
    &= \pm 2\iu\sqrt{\frac{(2\l_1+1)(2\l_2+1)(2\l_3+1)}{4\pi}}
    \notag \\
    &\qquad\times \Wjm{\l_1}{\l_2}{\l_3}{\pm2}{0}{\mp 2} 
    \,. 
}
%--------------------------------------------------%
The last term in the above denotes the Wigner $3j$-symbol. The off-diagonal elements of the covariance induced by the anisotropies of the rotation angle are given by
%--------------------------------------------------%
\al{
    \ave{E'_{\l m}B'_{\l'm'}}_{\rm CMB} 
    &= \sum_{x={\rm rei,rec}}\sum_{LM}\Wjm{\l}{\l'}{L}{m}{m'}{M} f^{\alpha,x}_{\l L \l'} (\alpha^x_{LM})^* 
    \,, \label{Eq:weight} 
}
%--------------------------------------------------%
where $\l\not=\l'$, $m\not=-m'$, and the operator $\ave{\cdots}_{\rm CMB}$ denotes an ensemble average over the realizations of CMB and noise with a fixed realization of $\alpha_x(\hatn)$. We ignore the correlation between recombination and reionization $E$-modes since such correlations are small \cite{Nakatsuka:2022}. The weight function is 
%--------------------------------------------------%
\al{
    f^{\alpha,x}_{\l L \l'} &= -\iu\frac{W^+_{\l'L\l}-W^-_{\l'L\l}}{2}\tCEE_{\l,x}
    \,, \label{Eq:weight:a}
}
%--------------------------------------------------%
where $\tCEE_{\l,x}$ is the cross-power spectrum between the reionization and total $E$-modes. We assume that $\tCEE_{\l,{\rm rei}}$ is equivalent to the total $E$-mode power spectrum, $\tCEE_\l$, at $2\leq\l\leq 10$ and zero otherwise. The term originating from the lensing $B$ mode is ignored since the improvement of the sensitivity to $\alpha$ by the inclusion of this term is negligible \cite{PB15:rot}. 

From Eq.~\eqref{Eq:weight}, the unnormalized quadratic estimator of $\alpha^{\rm rei}$ is constructed as a convolution of the $E$ and $B$ modes with the weight function of \eq{Eq:weight:a}: 
%--------------------------------------------------%
\al{
    (\bar{\alpha}^{\rm rei}_{LM})^* = \sum_{\l\l'mm'}\Wjm{\l}{\l'}{L}{m}{m'}{M}f^{\alpha,\rm rei}_{\l L \l'}\ol{E}_{\l m}\ol{B}_{\l'm'}
    \,. \label{Eq:uest}
}
%--------------------------------------------------%
Here, $\ol{E}_{\l m}$ and $\ol{B}_{\l m}$ are the observed multipoles filtered by their inverse variance. For a diagonal filtering, we obtain $\ol{X}_{\l m}=\hX_{\l m}/\hCXX_\l$, where $X$ is either $E$ or $B$ and $\hCXX_\l$ is the power spectrum of the observed multipoles, $\hX_{\l m}$. In practice, however, the diagonal filtering is not optimal. In this paper, we use the optimal filtering as detailed in Eq.~(2.11) of Ref.~\cite{Lonappan:LiteBIRD-lens}. Finally, we correct for the mean-field bias, $\ave{\bar{\alpha}^{\rm rei}_{LM}}$, and normalize to obtain the rotation angle: 
%--------------------------------------------------%
\al{
    \hat{\alpha}^{\rm rei}_{LM} = A_L(\bar{\alpha}^{\rm rei}_{LM}-\ave{\bar{\alpha}^{\rm rei}_{LM}}) 
    \,. \label{Eq:aest}
}
%--------------------------------------------------%
The normalization $A_L$ is given by
%--------------------------------------------------%
\al{
    A_L = \bigg[\frac{1}{2L+1}\sum_{\l\l'}\frac{|f^{\alpha,\rm rei}_{\l L\l'}|^2}{\hCEE_\l\hCBB_{\l'}}\bigg]^{-1}
    \,. \label{Eq:Rec:N0}
}
%--------------------------------------------------%
We can compute \eq{Eq:uest,Eq:Rec:N0} with a computationally efficient method as described in Appendix A of \cite{Namikawa:2020:biref}. The entire code for reconstructing cosmic birefringence on the full sky is available at \url{https://github.com/toshiyan/cmblensplus/tree/master}.

\subsection{Anisotropic cosmic birefringence from multiple redshifts}

We now extend the previous discussion to the measurement of anisotropic cosmic birefringence across multiple redshifts. To do this, we approximate the total observed polarization map as a sum of the contributions from each redshift: 
%--------------------------------------------------%
\al{
    Q'(\hatn)\pm\iu U'(\hatn) = \sum_{z}[Q_z(\hatn)\pm\iu U_z(\hatn)]\E^{\pm 2\iu \alpha_z(\hatn)}
    \,. \label{Eq:qurot:z}
}
%--------------------------------------------------%
From this, we derive the $E$- and $B$-modes, and the off-diagonal elements of the covariance are given by
%--------------------------------------------------%
\al{
    \ave{E'_{\l m}B'_{\l'm'}}_{\rm CMB} 
    &= \sum_{z}\sum_{LM}\Wjm{\l}{\l'}{L}{m}{m'}{M} f^{\alpha,z}_{\l L \l'} (\alpha^z_{LM})^* 
    \,. \label{Eq:weight:z} 
}
%--------------------------------------------------%
The weight function in this expression is
%--------------------------------------------------%
\al{
    f^{\alpha,z}_{\l L \l'} &= -\iu\frac{W^+_{\l'L\l}-W^-_{\l'L\l}}{2}\tCEE_{\l,z}
    \,, \label{Eq:weight:a:z}
}
%--------------------------------------------------%
where $\tCEE_{\l,z}$ represents the cross-power spectrum between $E$-mode generated at redshift $z$ and the total $E$-mode. 

Analogous to the estimator for $\alpha^{\rm rei}$, we define the estimator for $\alpha^z$ using Eq.~\eqref{Eq:weight:z} as follows: 
%--------------------------------------------------%
\al{
    (\hat{\alpha}^{z,{\rm bias}}_{LM})^* = A_L^{z}\sum_{\l\l'mm'}\Wjm{\l}{\l'}{L}{m}{m'}{M}f^{\alpha,z}_{\l L \l'}\ol{E}_{\l m}\ol{B}_{\l'm'}
    \,. \label{Eq:quest:z}
}
%--------------------------------------------------%
Here, $A_L^z$ is the same as the normalization defined in Eq.~\eqref{Eq:Rec:N0} but with $f^{\alpha,z}$ instead of $f^{\alpha,{\rm rei}}$. However, this estimator is biased because its expectation value includes contributions from anisotropic birefringence at different redshifts: 
\al{
    \ave{\hat{\alpha}^{z,{\rm bias}}_{LM}} = \sum_{z'}R_L^{zz'}\alpha^{z'}_{LM} \equiv \bR{R}_L \bm{\alpha}_{LM}
    \,. 
}
Here, we define the response function:
\al{
    R_L^{zz'} = A_L^{z} \bigg[\frac{1}{2L+1}\sum_{\l\l'}\frac{f^{\alpha,z}_{\l L\l'}f^{\alpha,z'}_{\l L\l'}}{\hCEE_\l\hCBB_{\l'}}\bigg]^{-1}
    \,. 
}
Following the approach used in constructing the estimator for CMB lensing analysis \cite{Namikawa:2012:bhe,Namikawa:2013:bhe-pol}, we define the unbiased quadratic estimator for the anisotropic cosmic birefringence at each redshift as
\al{
    \hat{\alpha}^z_{LM} = \bR{R}^{-1}_L\bm{\hat{\alpha}}^{{\rm bias}}_{LM} = \sum_{z'}\{\bR{R}_L^{-1}\}_{zz'} \hat{\alpha}^{z',{\rm bias}}_{LM}
    \,.
}
This estimator is both unbiased and an optimal quadratic estimator, satisfying the necessary conditions for accurate anisotropic birefringence measurements at different redshifts.

Arcari et al. \cite{Arcari:2024nhw} recently estimated the signal-to-noise ratio (SNR) of the cross-power spectrum between anisotropic cosmic birefringence and galaxy number density fluctuations. In their analysis, the reconstruction noise was calculated under the assumption that the source redshift of CMB photons corresponds to the epoch of recombination. However, this assumption is not appropriate for their specific purpose. Since the cross-power spectrum is primarily influenced by correlations in the late-time universe, it is necessary to reconstruct anisotropic cosmic birefringence at lower redshifts. To accurately account for this, the appropriate reconstruction noise should be derived using the estimator designed for low-redshift birefringence, as discussed earlier. Given that the SNR for low-redshift anisotropic cosmic birefringence is significantly smaller than that at recombination, their original estimate of the SNR for the cross-power spectrum would likely be reduced.

%////////////////////////////////////////%
\section{Data and simulations} \label{sec:datasim}
%////////////////////////////////////////%

%<><><><><><><><><><>%
\begin{figure}[t]
\bc
\includegraphics[width=8.5cm]{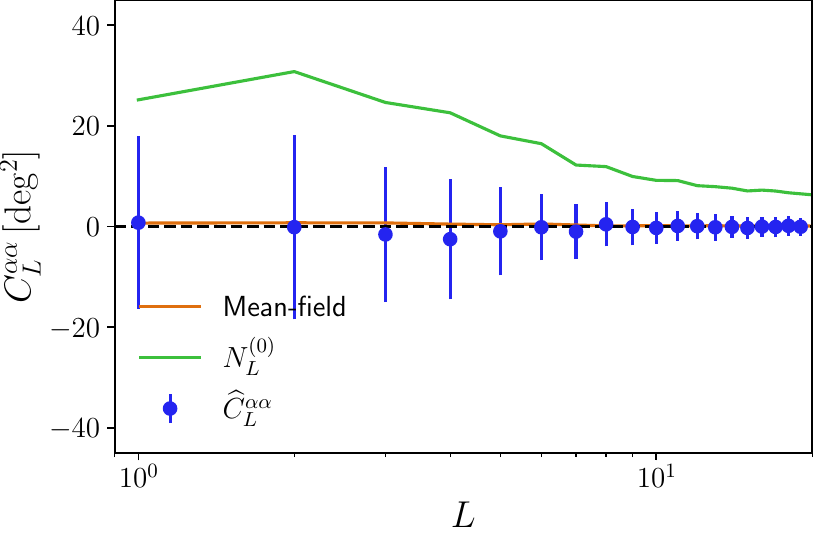}
\caption{
Angular power spectrum of anisotropic cosmic birefringence at reionization measured from the Planck PR4 simulation with the $70\%$ Galactic mask, shown in blue. The CMB polarization and noise are derived from the Commander component separation method. The error bars denote the standard deviation with $400$ realizations. We also show the mean-field bias (orange) and disconnected bias, $N^{(0)}$ (green), which corresponds to the noise power spectrum of $\alpha^{\rm rei}_{LM}$. 
}
\label{Fig:aps-sim-null}
\ec
\end{figure}
%<><><><><><><><><><>%

%<><><><><><><><><><>%
\begin{figure}[t]
\bc
\includegraphics[width=8.5cm]{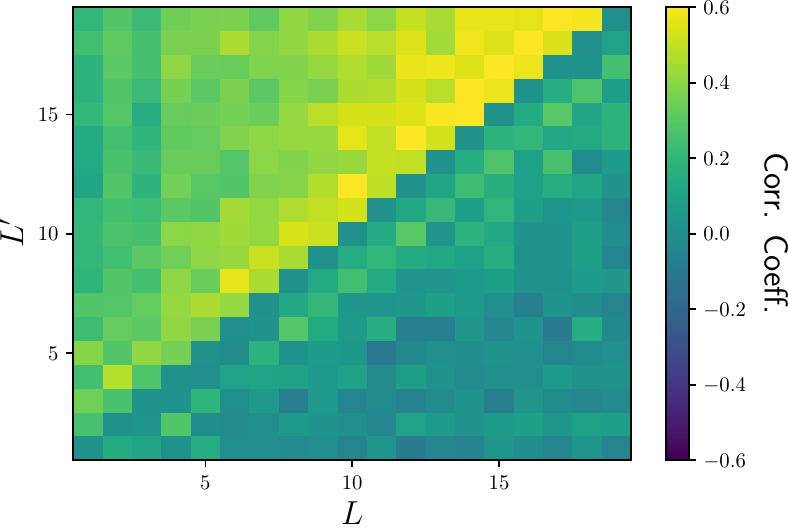}
\caption{
Correlation coefficients of the angular power spectrum of anisotropic cosmic birefringence measured from the Planck PR4 simulation, as shown in Fig.~\ref{Fig:aps-sim-null}. The lower-left triangle displays the results when the realization-dependent disconnected bias defined in Eq.(27) of Ref.~\cite{Namikawa:2016:biref-est} is used for estimating the disconnected bias, while the upper-left triangle shows, for comparison, the case where the simulation average of the disconnected bias defined in Eq.(26) of Ref.~\cite{Namikawa:2016:biref-est} is used.
}
\label{Fig:aps-sim-null_corr}
\ec
\end{figure}
%<><><><><><><><><><>%

%<><><><><><><><><><>%
\begin{figure}[t]
\bc
\includegraphics[width=8.5cm]{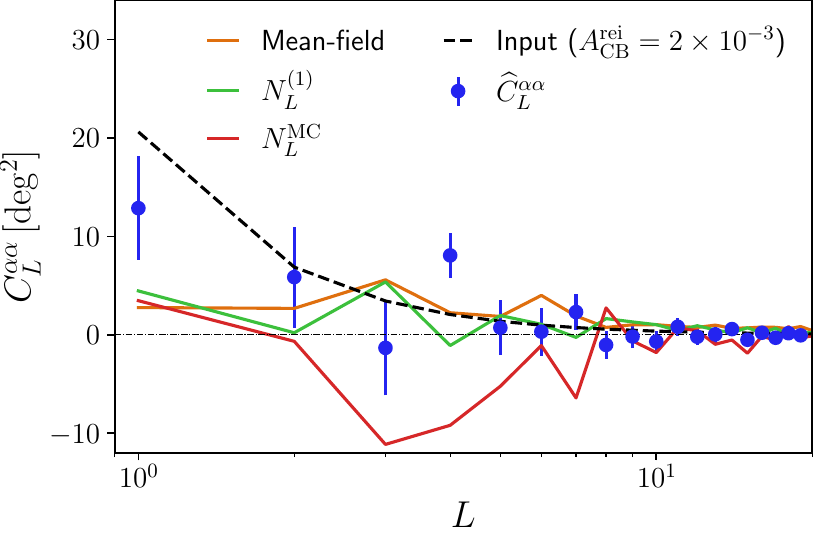}
\caption{
Angular power spectrum of anisotropic cosmic birefringence obtained from $400$ simulation realizations, $\hat{C}_L^{\alpha\alpha}$ (blue points). The error bars represent the standard deviation divided by $\sqrt{400}$. The simulated CMB polarization is rotated by a scale-invariant anisotropic cosmic birefringence with $A_{\rm CB}^{\rm rei}=2\times10^{-3}$, applied to the reionization signals. The mean-field bias (orange), 
N1 bias (green), and other biases (red) are also shown for comparison.
}
\label{Fig:aps-sim-acb}
\ec
\end{figure}
%<><><><><><><><><><>%

% Data
We use the Planck PR4 polarization data to reconstruct anisotropic cosmic birefringence, employing the curved-sky quadratic estimator defined in Eq.~\eqref{Eq:aest} to extract large-scale birefringence anisotropies. 

% simulation
The Planck PR4 polarization map is produced using two different component separation algorithms: Commander and SEVEM. We use both maps and the corresponding simulation sets to reconstruct anisotropic cosmic birefringence. Additionally, we generate a simulation where the CMB signal components are replaced with a signal that includes nonzero anisotropic cosmic birefringence, whose power spectrum follows a scale-invariant model. In this simulation, we first generate $\alpha_{LM}^{\rm rei}$ from a scale-invariant spectrum $C_L^{\alpha\alpha}=A_{\rm CB}^{\rm rei}\times 2\pi/L(L+1)$, and then rotate the CMB polarization signals at $\l\leq 10$ with this rotation field. We assume $A_{\rm CB}^{\rm rei}=2\times10^{-3}$.

% quadratic estimator
For our analysis, the polarization map uses a Galactic mask, and therefore, the diagonal filtering described earlier is not optimal. To address this, we obtain $\ol{E}_{\l m}$ and $\ol{B}_{\l m}$ through a more optimal method, as detailed in Eq.~(2.11) of Ref.~\cite{Lonappan:LiteBIRD-lens}. In this calculation, we use an estimate of the noise power spectrum for the noise covariance matrix, which is derived from the auto- and cross-power spectra of the detector-split data. In the quadratic estimator, the $B$-mode multipoles at $2\leq\l'\leq128$ are used. The estimator automatically removes the $E$-mode multipole at $\l\geq 11$ as $f_{\ell L\ell'}^{\alpha,{\rm rei}}$ contains $\tCEE_{\l,{\rm rei}}$, leaving us to use the $E$-mode multipole at $2\leq\l\leq 10$. 

% Power spectrum calculation
We evaluate the mean-field bias by averaging over the standard simulation realizations. Note that mean-field bias is also induced by the miscalibrated polarization angle \cite{Namikawa:2020:biref}, which is not included in the simulations provided by the Planck Collaboration. We test this effect by rotating the polarization map in the simulation as we discuss in the next section.

% Normalization correction
In practice, partial sky coverage introduces mixing between different multipoles, and the diagonal normalization does not provide an accurate normalization. To correct for this, we follow the approach used in CMB lensing analyses. We first reconstruct $\alpha$ from a masked simulation that includes nonzero anisotropic birefringence. Next, we cross-correlate the input $\alpha$ with the reconstructed $\alpha$ and rescale the normalization at each $L$ so that the cross-power spectrum agrees with the input birefringence power spectrum. 

% Power spectrum estimation
From the reconstructed $\alpha$, we estimate the cosmic birefringence spectrum following Ref.~\cite{Namikawa:2016:biref-est}. The power spectrum of the estimator defined in \eq{Eq:aest} is a four-point function and is decomposed into the disconnected and connected parts. The disconnected part, $N_L^{(0)}$, is nonzero even in the absence of birefringence. We evaluate the disconnected bias using the realization-dependent approach as defined in Eq.~(27) of Ref.~\cite{Namikawa:2016:biref-est} and subtract it from the power spectrum of the estimator. The N1 bias related to the connected part of the trispectrum is estimated using the simulations with nonzero birefringence. We also consider the residual power spectrum in the PR4 standard simulation as additional bias from the connected four-point function, $N^{\rm MC}_L$, arising from lensing \cite{Namikawa:2016:biref-est}, partial-sky coverage, and noise. We subtract all these terms from the power spectrum of the estimator to extract the power spectrum of the anisotropic cosmic birefringence signal as described in Eq.~(33) of Ref.~\cite{Namikawa:2016:biref-est}. 

Figure~\ref{Fig:aps-sim-null} shows the angular power spectrum of anisotropic cosmic birefringence from the PR4 simulation, which contains no birefringence signals. We use the $70\%$ Galactic mask and simulation set with the Commander component separation method. We do not apply any normalization correction to the estimator. The estimated power spectrum is consistent with a null result. Figure~\ref{Fig:aps-sim-null_corr} plots the correlation coefficients of the angular power spectrum. The realization-dependent method for $N_L^{(0)}$ successfully reduces the off-diagonal covariance. Figure~\ref{Fig:aps-sim-acb} shows the angular power spectrum obtained from a simulation that includes anisotropic cosmic birefringence signals. We employ the $70\%$ Galactic mask and simulation set with the Commander component separation method. The power spectrum estimator accurately reproduces the input angular power spectrum. The mean-field, $N^{(1)}$, and $N^{\rm MC}$ are smaller than or comparable to the input signal with $A_{\rm CB}^{\rm rei}=2\times10^{-3}$ which is much lower than the value constrained from the real data as we show in the next section. Note that $N^{(0)}$ is much larger than the other spectra and becomes approximately $100$ [deg$^2$] at $L<10$ and $\mC{O}(10)$ [deg$^2$] at $10\leq L\leq 100$. This large value of $N^{(0)}$ arises from the normalization correction applied to the estimator. It reflects a significant suppression of reconstructed anisotropic birefringence signals on large angular scales.

%////////////////////////////////////////%
\section{Results} \label{sec:results}
%////////////////////////////////////////%

%<><><><><><><><><><>%
\begin{figure*}[th]
\bc
\includegraphics[width=8.5cm]{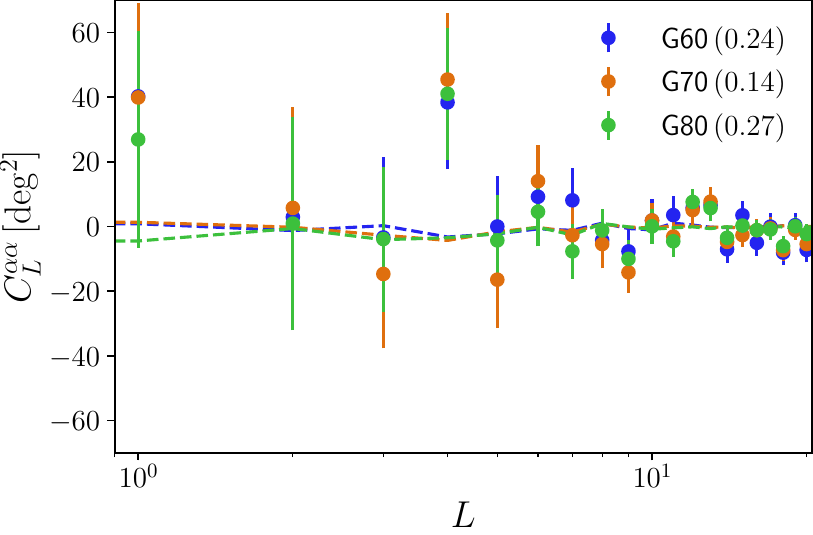}
\includegraphics[width=8.5cm]{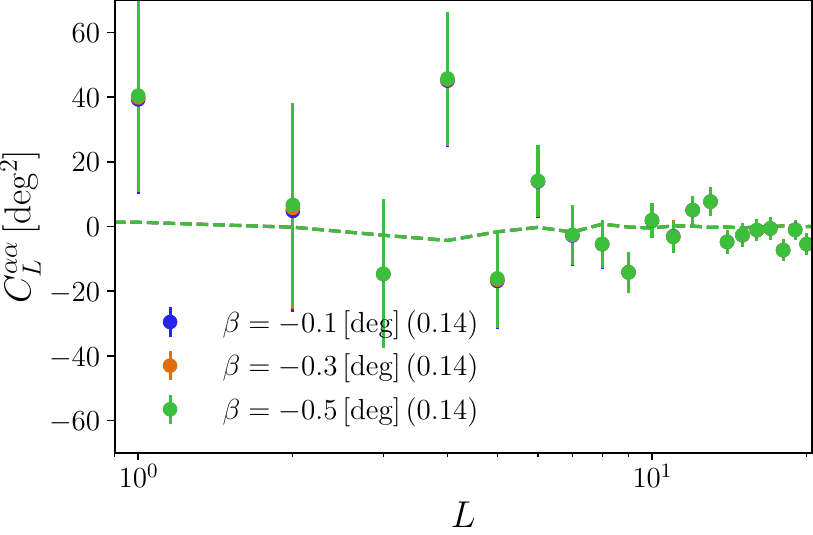}
\includegraphics[width=8.5cm]{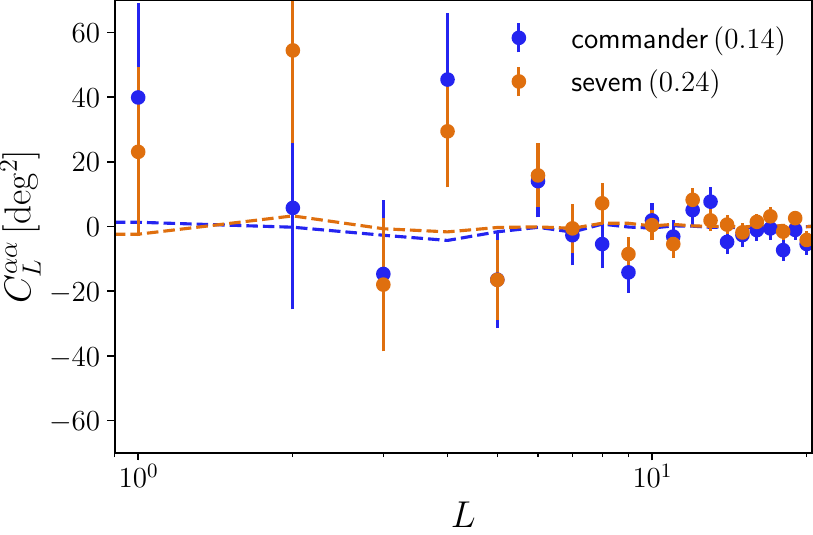}
\includegraphics[width=8.5cm]{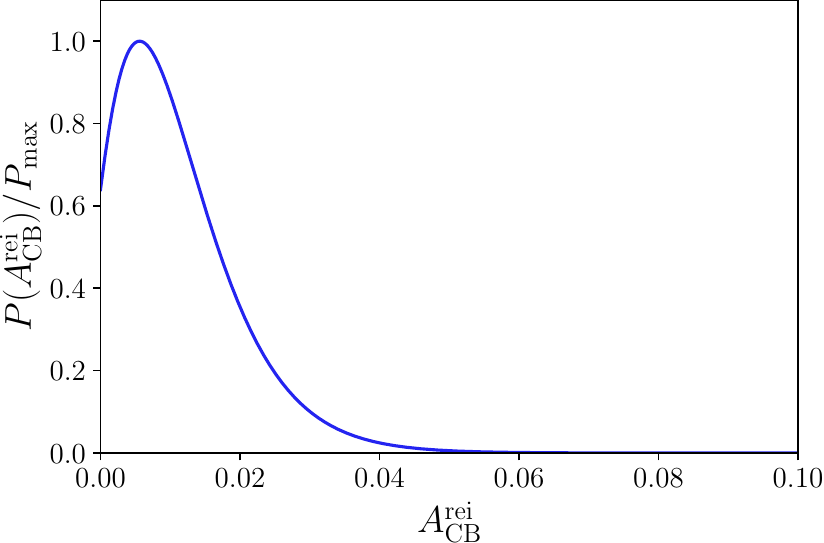}
\caption{
Angular power spectrum of anisotropic cosmic birefringence measured from the PR4 polarization data over nearly the full sky. The panels show results for varying the Planck Galactic mask (top left), applying a uniform rotation by $\beta$ to the $Q$ and $U$ maps before reconstruction (top right), and using different component separation methods (Commander and SEVEM, bottom left). The error bars represent the statistical uncertainties, estimated from standard $\Lambda$CDM simulations. The dashed lines indicate the mean-field bias, evaluated from simulations. The values in parentheses next to the labels represent the $\chi^2$-PTE. The bottom-right panel shows the constraint on $A_{\rm CB}^{\rm rei}$ derived from the measured birefringence power spectrum in the baseline case.
}
\label{Fig:main}
\ec
\end{figure*}
%<><><><><><><><><><>%

Figure~\ref{Fig:main} shows the cosmic birefringence power spectrum measured from the Planck PR4 polarization data, with error bars derived from the standard simulations. In our baseline case, we employ the polarization map obtained from the Commander method for Galactic foreground cleaning and use the Planck $70\%$ Galactic mask. At large-angular scales, a nonzero isotropic birefringence can introduce a mean-field bias \cite{Namikawa:2020:biref}. To mitigate the impact of a potential isotropic birefringence angle $\beta$, we apply a uniform derotation of $-\beta$ to the polarization map prior to reconstruction. We do not find any relevant difference in the measured birefringence spectrum for different values of $\beta=-0.1/0.3/0.5$ degrees. At large angular scales, the Galactic foreground can also potentially affect the measurement. Therefore, we also measure the angular power spectrum by varying the Galactic mask and the foreground cleaning method. In all cases, we compute the $\chi^2$-PTE of the measured power spectrum for multipoles in the range $1\leq L\leq 20$, finding that the measured spectrum is consistent with a null result.

Using the measured power spectrum in the baseline case, $\hat{C}_L^{\alpha\alpha}$, we constrain the amplitude of the scale-invariant power spectrum, $A_{\rm CB}^{\rm rei}$. Following the previous studies \cite{Namikawa:2020:biref,SPT:2020:biref}, we evaluate the constraint on $A_{\rm CB}^{\rm rei}$ using the following likelihood, motivated by the likelihood approximation from Ref.~\cite{Hamimeche:2008ai}: 
\al{
    -2\ln\mC{L}(A_{\rm CB}^{\rm rei}) = \sum_{L,L'}g(x_L)C_L^{\rm f}\bR{C}^{-1}_{LL'}C_{L'}^{\rm f}g(x_{L'}) 
    \,. 
}
where $g(x)={\rm sign}(x-1)\sqrt{2(x-\ln x-1)}$ and $x_L=C_L^{\hat{\alpha}\hat{\alpha},\rm obs}/C_L^{\hat{\alpha}\hat{\alpha},\rm th}$ with
\al{
    C_L^{\hat{\alpha}\hat{\alpha},\rm obs} &= \hat{C}_L^{\alpha\alpha}+N_L^{(0)}+N_L^{\rm MC}
    \,, \\
    C_L^{\hat{\alpha}\hat{\alpha},\rm th} &= A_{\rm CB}^{\rm rei}\left(C_L^{\alpha\alpha,{\rm fid}}+N_L^{(1)}\right)+N_L^{(0)}+N_L^{\rm MC}
    \,. 
}
Here, $C_L^{\alpha\alpha,{\rm fid}}=2\pi/L(L+1)$, $C_L^{\rm f}$ corresponds to $C_L^{\hat{\alpha}\hat{\alpha},\rm th}$ evaluated at $A_{\rm CB}^{\rm rei}=0$, and ${\rm C}_{LL'}$ is the covariance matrix of $C_L^{\rm f}$. We show $\mC{L}$ as a function of $A_{\rm CB}^{\rm rei}$ in Fig.~\ref{Fig:main}. From this analysis, we find an upper limit of $A_{\rm CB}^{\rm rei}<1.3\times 10^{-2}$ at the $1\,\sigma$ level. It is important to note that the model proposed by Ref.~\cite{Ferreira:2023}, which predicts $A_{\rm CB}^{\rm rei}\simeq 10^{-5}$, is still consistent with current observational constraints. However, this model would be excluded if we were to naively apply the anisotropic cosmic birefringence power spectrum measured in the previous studies \cite{Gluscevic:2012,PB15:rot,BICEP2:2017lpa,Contreras:2017,Namikawa:2020:biref,SPT:2020:biref,Gruppuso:2020,Bortolami:2022whx,BK-LoS:2023,Zagatti:2024}.

%////////////////////////////////////////%
\section{Summary} \label{sec:summary}
%////////////////////////////////////////%

%<><><><><><><><><><>%
\begin{figure}[t]
\bc
\includegraphics[width=8cm]{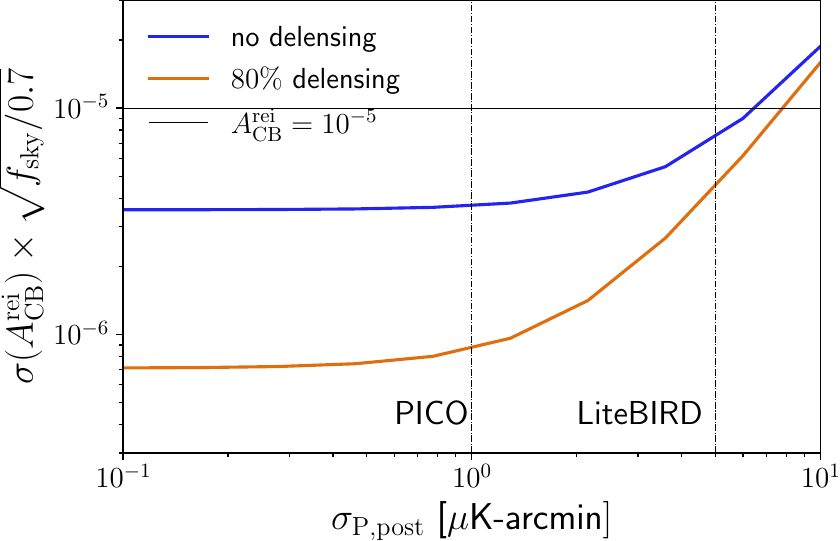}
\caption{
Uncertainty on the amplitude of the scale-invariant power spectrum, $A_{\rm CB}^{\rm rei}$, derived from the cosmic birefringence power spectrum, shown as a function of polarization noise after Galactic foreground cleaning. We present results for both the delensed and non-delensed cases. The black horizontal line represents the prediction from Ref.~\cite{Ferreira:2023}, while the vertical lines indicate the expected noise levels for the PICO and LiteBIRD experiments.
}
\label{Fig:forecast}
\ec
\end{figure}
%<><><><><><><><><><>%

We explored tomographic constraints on anisotropic cosmic birefringence using CMB polarization data. Specifically, we measured the angular power spectrum of anisotropic cosmic birefringence generated during the reionization epoch, using the Planck PR4 polarization maps. Our analysis found the power spectrum to be consistent with null, leading to an upper bound on the amplitude of the scale-invariant power spectrum: $A_{\rm CB}^{\rm rei}< 1.3\times 10^{-2}\,(1\,\sigma)$. 

% Reionization uncertainty
This measurement relies on the reionization signal at large angular scales. Although the reionization history is constrained by the $E$-mode power spectrum, uncertainties remain in the large-scale $E$-mode power spectrum from Planck PR4 \cite{Tristram:2023haj}. These uncertainties in the reionization history could potentially introduce a bias in the estimator of anisotropic cosmic birefringence. However, since the anisotropic cosmic birefringence is a higher-order correlation function, it is significantly less sensitive to such biases compared to the $E$-mode power spectrum. Therefore, any bias introduced by uncertainties in the reionization history would be negligible relative to the measurement error in the anisotropic cosmic birefringence estimation.

% future
Measurements of anisotropic cosmic birefringence are crucial for testing new physical theories, particularly those involving parity violation. Future CMB experiments such as LiteBIRD \cite{LiteBIRD} and PICO \cite{NASAPICO:2019thw} will offer more precise measurements of anisotropic cosmic birefringence generated during the reionization epoch. Figure~\ref{Fig:forecast} illustrates the expected constraints on the amplitude of the scale-invariant power spectrum as a function of CMB experiment noise levels. We compute the expected uncertainty on $\sigma(A_{\rm CB}^{\rm rei})$ using the following expression: 
\al{
    \sigma(A_{\rm CB}^{\rm rei}) = \left[\sum_L\frac{2L+1}{2}f_{\rm sky}\left(\frac{C_L^{\alpha\alpha,\rm fid}}{A_L}\right)^2\right]^{-1/2}
    \,. 
}
where $C_L^{\alpha\alpha,\rm fid}=2\pi/L(L+1)$ and $f_{\rm sky}$ is the fraction of observed sky. The normalization given in Eq.~\eqref{Eq:Rec:N0}, $A_L$, corresponds to the reconstruction noise power spectrum. The observed $E$- and $B$-mode power spectra are modeled as the sum of the lensed CMB and white-noise power spectra. We assume that, after Galactic foreground cleaning, the noise level is characterized by white noise with an amplitude $\sigma_{\rm P,post}$. Additionally, we account for the delensing of $B$-mode polarization, which reduces the cosmic variance from lensing and enhances sensitivity to primary $B$-modes \cite{Kesden:2002ku,Seljak:2003pn}. Specifically, we consider an $80\%$ delensing scenario for the space-based experiments, achievable through a combination of future ground-based CMB experiments and external mass tracers \cite{CMBS4,Namikawa:2023:LB-delens}. In this scenario, the lensing $B$-mode power spectrum is scaled by a factor of $0.2$ before calculating $A_L$ in Eq.~\eqref{Eq:Rec:N0}. LiteBIRD and PICO will improve the sensitivity to $A_{\rm CB}^{\rm rei}$ by more than three orders of magnitude compared to Planck, enabling them to test the model proposed by Ref.~\cite{Ferreira:2023}.

%////////////////////////////////////////%
% BACK MATTER 
%////////////////////////////////////////%

% Appendix %
\appendix

\begin{acknowledgments}
We thank Eiichiro Komatsu and Ippei Obata for their useful comments and discussion. This work is supported in part by JSPS KAKENHI Grant No. JP20H05859 and No. JP22K03682. Part of this work uses resources of the National Energy Research Scientific Computing Center (NERSC). The Kavli IPMU is supported by World Premier International Research Center Initiative (WPI Initiative), MEXT, Japan. 
\end{acknowledgments}

% References %
%\bibliographystyle{apsrev}
\bibliographystyle{mybst}
\bibliography{cite}

\providecommand{\href}[2]{#2}\begingroup\raggedright\begin{thebibliography}{10}

\bibitem{Minami:2020:biref}
Y.~Minami and E.~Komatsu {\em \prl} {\bf 125} (2020), no.~22 221301, \href{http://arxiv.org/abs/2011.11254}{{\tt arXiv:2011.11254}}.

\bibitem{Diego-Palazuelos:2022}
P.~Diego-Palazuelos {\em et~al.} {\em \prl} {\bf 128} (2022), no.~9 091302, \href{http://arxiv.org/abs/2201.07682}{{\tt arXiv:2201.07682}}.

\bibitem{Eskilt:2022:biref-freq}
J.~R. Eskilt {\em \aap} {\bf 662} (2022) A10, \href{http://arxiv.org/abs/2201.13347}{{\tt arXiv:2201.13347}}.

\bibitem{Eskilt:2022:biref-const}
J.~R. Eskilt and E.~Komatsu {\em \prd} {\bf 106} (2022), no.~6 063503, \href{http://arxiv.org/abs/2205.13962}{{\tt arXiv:2205.13962}}.

\bibitem{Eskilt:2023:EDE}
J.~R. Eskilt, L.~Herold, E.~Komatsu, K.~Murai, T.~Namikawa, and F.~Naokawa \href{http://arxiv.org/abs/2303.15369}{{\tt arXiv:2303.15369}}.

\bibitem{Komatsu:2022:review}
E.~Komatsu {\em Nature Rev. Phys.} {\bf 4} (2022), no.~7 452--469, \href{http://arxiv.org/abs/2202.13919}{{\tt arXiv:2202.13919}}.

\bibitem{Nakai:2023}
Y.~Nakai, R.~Namba, I.~Obata, Y.-C. Qiu, and R.~Saito {\em JHEP} {\bf 01} (2024) 057, \href{http://arxiv.org/abs/2310.09152}{{\tt arXiv:2310.09152}}.

\bibitem{Carroll:1998:DE}
S.~M. Carroll {\em \prl} {\bf 81} (1998) 3067--3070, \href{http://arxiv.org/abs/astro-ph/9806099}{{\tt astro-ph/9806099}}.

\bibitem{Liu:2006:biref-time-evolve}
G.-C. Liu, S.~Lee, and K.-W. Ng {\em \prl} {\bf 97} (2006) 161303, \href{http://arxiv.org/abs/astro-ph/0606248}{{\tt astro-ph/0606248}}.

\bibitem{Panda:2010}
S.~Panda, Y.~Sumitomo, and S.~P. Trivedi {\em \prd} {\bf 83} (2011) 083506, \href{http://arxiv.org/abs/1011.5877}{{\tt arXiv:1011.5877}}.

\bibitem{Fujita:2020aqt}
T.~Fujita, Y.~Minami, K.~Murai, and H.~Nakatsuka {\em Phys. Rev. D} {\bf 103} (2021), no.~6 063508, \href{http://arxiv.org/abs/2008.02473}{{\tt arXiv:2008.02473}}.

\bibitem{Fujita:2020ecn}
T.~Fujita, K.~Murai, H.~Nakatsuka, and S.~Tsujikawa {\em Phys. Rev. D} {\bf 103} (2021), no.~4 043509, \href{http://arxiv.org/abs/2011.11894}{{\tt arXiv:2011.11894}}.

\bibitem{Choi:2021aze}
G.~Choi, W.~Lin, L.~Visinelli, and T.~T. Yanagida {\em Phys. Rev. D} {\bf 104} (2021), no.~10 L101302, \href{http://arxiv.org/abs/2106.12602}{{\tt arXiv:2106.12602}}.

\bibitem{Obata:2021}
I.~Obata {\em \jcap} {\bf 09} (2022) 062, \href{http://arxiv.org/abs/2108.02150}{{\tt arXiv:2108.02150}}.

\bibitem{Gasparotto:2022uqo}
S.~Gasparotto and I.~Obata {\em JCAP} {\bf 08} (2022), no.~08 025, \href{http://arxiv.org/abs/2203.09409}{{\tt arXiv:2203.09409}}.

\bibitem{Galaverni:2023}
M.~Galaverni, F.~Finelli, and D.~Paoletti {\em \prd} {\bf 107} (2023), no.~8 083529, \href{http://arxiv.org/abs/2301.07971}{{\tt arXiv:2301.07971}}.

\bibitem{Murai:2022:EDE}
K.~Murai, F.~Naokawa, T.~Namikawa, and E.~Komatsu {\em \prd} {\bf 107} (2023) L041302, \href{http://arxiv.org/abs/2209.07804}{{\tt arXiv:2209.07804}}.

\bibitem{Kochappan:2024:biref}
J.~Kochappan, L.~Yin, B.-H. Lee, and T.~Ghosh \href{http://arxiv.org/abs/2408.09521}{{\tt arXiv:2408.09521}}.

\bibitem{Finelli:2009}
F.~Finelli and M.~Galaverni {\em \prd} {\bf 79} (2009) 063002, \href{http://arxiv.org/abs/0802.4210}{{\tt arXiv:0802.4210}}.

\bibitem{Sigl:2018:biref-sup}
G.~Sigl and P.~Trivedi {\em {}} (11, 2018) \href{http://arxiv.org/abs/1811.07873}{{\tt arXiv:1811.07873}}.

\bibitem{Liu:2016dcg}
G.-C. Liu and K.-W. Ng {\em Phys. Dark Univ.} {\bf 16} (2017) 22--25, \href{http://arxiv.org/abs/1612.02104}{{\tt arXiv:1612.02104}}.

\bibitem{Fedderke:2019:biref}
M.~A. Fedderke, P.~W. Graham, and S.~Rajendran {\em \prd} {\bf 100} (2019) 015040, \href{http://arxiv.org/abs/1903.02666}{{\tt arXiv:1903.02666}}.

\bibitem{Zhang:2024dmi}
D.~Zhang, E.~G.~M. Ferreira, I.~Obata, and T.~Namikawa \href{http://arxiv.org/abs/2408.08063}{{\tt arXiv:2408.08063}}.

\bibitem{Takahashi:2020tqv}
F.~Takahashi and W.~Yin {\em JCAP} {\bf 04} (2021) 007, \href{http://arxiv.org/abs/2012.11576}{{\tt arXiv:2012.11576}}.

\bibitem{Kitajima:2022jzz}
N.~Kitajima, F.~Kozai, F.~Takahashi, and W.~Yin {\em JCAP} {\bf 10} (2022) 043, \href{http://arxiv.org/abs/2205.05083}{{\tt arXiv:2205.05083}}.

\bibitem{Jain:2022jrp}
M.~Jain, R.~Hagimoto, A.~J. Long, and M.~A. Amin {\em JCAP} {\bf 10} (2022) 090, \href{http://arxiv.org/abs/2208.08391}{{\tt arXiv:2208.08391}}.

\bibitem{Gonzalez:2022mcx}
D.~Gonzalez, N.~Kitajima, F.~Takahashi, and W.~Yin \href{http://arxiv.org/abs/2211.06849}{{\tt arXiv:2211.06849}}.

\bibitem{Myers:2003fd}
R.~C. Myers and M.~Pospelov {\em \prl} {\bf 90} (2003) 211601, \href{http://arxiv.org/abs/hep-ph/0301124}{{\tt hep-ph/0301124}}.

\bibitem{Balaji:2003sw}
K.~R.~S. Balaji, R.~H. Brandenberger, and D.~A. Easson {\em \jcap} {\bf 12} (2003) 008, \href{http://arxiv.org/abs/hep-ph/0310368}{{\tt hep-ph/0310368}}.

\bibitem{Arvanitaki:2009fg}
A.~Arvanitaki, S.~Dimopoulos, S.~Dubovsky, N.~Kaloper, and J.~March-Russell {\em \prd} {\bf 81} (2010) 123530, \href{http://arxiv.org/abs/0905.4720}{{\tt arXiv:0905.4720}}.

\bibitem{Cornelison:2022:BICEP3}
J.~Cornelison {\em et~al.} {\em Proc. SPIE Int. Soc. Opt. Eng.} {\bf 12190} (2022) 121901X, \href{http://arxiv.org/abs/2207.14796}{{\tt arXiv:2207.14796}}.

\bibitem{BICEPArray}
L.~Moncelsi, P.~A.~R. Ade, Z.~Ahmed, M.~Amiri, D.~Barkats, R.~{Basu Thakur}, C.~A. Bischoff, J.~J. Bock, V.~Buza, J.~R. Cheshire, {\em et~al.} {\em Proc. SPIE Int. Soc. Opt. Eng.} {\bf 11453} (2020) 1145314, \href{http://arxiv.org/abs/2012.04047}{{\tt arXiv:2012.04047}}.

\bibitem{SimonsArray}
N.~{Stebor}, P.~{Ade}, Y.~{Akiba}, C.~{Aleman}, K.~{Arnold}, C.~{Baccigalupi}, B.~{Barch}, D.~{Barron}, S.~{Beckman}, A.~{Bender}, {\em et~al.}, {\it ``{The Simons Array CMB polarization experiment}''},  in {\em Millimeter, Submillimeter, and Far-Infrared Detectors and Instrumentation for Astronomy VIII} (W.~S. {Holland} and J.~{Zmuidzinas}, eds.), vol.~9914 of {\em Society of Photo-Optical Instrumentation Engineers (SPIE) Conference Series}, p.~99141H, July, 2016.

\bibitem{SimonsObservatory}
{The Simons Observatory Collaboration} {\em \jcap} {\bf 02} (2019) 056, \href{http://arxiv.org/abs/1808.07445}{{\tt arXiv:1808.07445}}.

\bibitem{CMBS4}
{CMB-S4 Collaboration} {\em \apj} {\bf 926} (2022), no.~1 54, \href{http://arxiv.org/abs/2008.12619}{{\tt arXiv:2008.12619}}.

\bibitem{LiteBIRD}
{LiteBIRD Collaboration} {\em \ptep} (2, 2022) \href{http://arxiv.org/abs/2202.02773}{{\tt arXiv:2202.02773}}.

\bibitem{Huffenberger:2019:dustEB}
K.~M. Huffenberger, A.~Rotti, and D.~C. Collins {\em Astrophys. J.} {\bf 899} (2020), no.~1 31, \href{http://arxiv.org/abs/1906.10052}{{\tt arXiv:1906.10052}}.

\bibitem{Cukierman:2022:dust}
A.~J. Cukierman, S.~E. Clark, and G.~Halal {\em Astrophys. J.} {\bf 946} (2023), no.~2 106, \href{http://arxiv.org/abs/2208.07382}{{\tt arXiv:2208.07382}}.

\bibitem{Hervias-Caimapo:2024}
C.~Herv\'\i{}as-Caimapo, A.~J. Cukierman, P.~Diego-Palazuelos, K.~M. Huffenberger, and S.~E. Clark \href{http://arxiv.org/abs/2408.06214}{{\tt arXiv:2408.06214}}.

\bibitem{SO:wiregrid}
M.~{Murata}, H.~{Nakata}, K.~{Iijima}, S.~{Adachi}, Y.~{Seino}, K.~{Kiuchi}, F.~{Matsuda}, M.~J. {Randall}, K.~{Arnold}, N.~{Galitzki}, B.~R. {Johnson}, B.~{Keating}, A.~{Kusaka}, J.~B. {Lloyd}, J.~{Seibert}, M.~{Silva-Feaver}, O.~{Tajima}, T.~{Terasaki}, and K.~{Yamada} {\em Review of Scientific Instruments} {\bf 94} (2023), no.~12 124502, \href{http://arxiv.org/abs/2309.02035}{{\tt arXiv:2309.02035}}.

\bibitem{Coppi:2022}
G.~{Coppi}, G.~{Conenna}, S.~{Savorgnano}, F.~{Carrero}, R.~{D{\"u}nner Planella}, N.~{Galitzki}, F.~{Nati}, and M.~{Zannoni}, {\it ``{PROTOCALC: an artificial calibrator source for CMB telescopes}''},  in {\em Millimeter, Submillimeter, and Far-Infrared Detectors and Instrumentation for Astronomy XI}, vol.~12190 of {\em Society of Photo-Optical Instrumentation Engineers (SPIE) Conference Series}, p.~1219015, 2022.
\newblock \href{http://arxiv.org/abs/2207.07595}{{\tt arXiv:2207.07595}}.

\bibitem{Ritacco:2023mgi}
A.~Ritacco {\em et~al.} {\em EPJ Web Conf.} {\bf 293} (2024) 00044, \href{http://arxiv.org/abs/2311.08307}{{\tt arXiv:2311.08307}}.

\bibitem{Lee:2013:const}
S.~Lee, G.-C. Liu, and K.-W. Ng {\em \prd} {\bf 89} (2014), no.~6 063010, \href{http://arxiv.org/abs/1307.6298}{{\tt arXiv:1307.6298}}.

\bibitem{Gubitosi:2014:biref-time}
G.~Gubitosi, M.~Martinelli, and L.~Pagano {\em \jcap} {\bf 12} (2014), no.~2014 020, \href{http://arxiv.org/abs/1410.1799}{{\tt arXiv:1410.1799}}.

\bibitem{Sherwin:2021:biref}
B.~D. Sherwin and T.~Namikawa {\em \mnras} {\bf 520} (2021) 3298–3304, \href{http://arxiv.org/abs/2108.09287}{{\tt arXiv:2108.09287}}.

\bibitem{Nakatsuka:2022}
H.~Nakatsuka, T.~Namikawa, and E.~Komatsu {\em \prd} {\bf 105} (2022) 123509, \href{http://arxiv.org/abs/2203.08560}{{\tt arXiv:2203.08560}}.

\bibitem{Naokawa:2023}
F.~Naokawa and T.~Namikawa {\em \prd} {\bf 108} (2023), no.~6 063525, \href{http://arxiv.org/abs/2305.13976}{{\tt arXiv:2305.13976}}.

\bibitem{Yin:2023:biref}
L.~Yin, J.~Kochappan, T.~Ghosh, and B.-H. Lee \href{http://arxiv.org/abs/2305.07937}{{\tt arXiv:2305.07937}}.

\bibitem{Naokawa:2024xhn}
F.~Naokawa, T.~Namikawa, K.~Murai, I.~Obata, and K.~Kamada \href{http://arxiv.org/abs/2405.15538}{{\tt arXiv:2405.15538}}.

\bibitem{Murai:2024yul}
K.~Murai \href{http://arxiv.org/abs/2407.14162}{{\tt arXiv:2407.14162}}.

\bibitem{Lee:2022:pSZ-biref}
N.~Lee, S.~C. Hotinli, and M.~Kamionkowski {\em \prd} {\bf 106} (2022), no.~8 083518, \href{http://arxiv.org/abs/2207.05687}{{\tt arXiv:2207.05687}}.

\bibitem{Namikawa:2023:pSZ}
T.~Namikawa and I.~Obata {\em \prd} {\bf 108} (2023), no.~8 083510, \href{http://arxiv.org/abs/2306.08875}{{\tt arXiv:2306.08875}}.

\bibitem{Carroll:1997:radio}
S.~M. Carroll and G.~B. Field {\em \prl} {\bf 79} (1997) 2394--2397, \href{http://arxiv.org/abs/astro-ph/9704263}{{\tt astro-ph/9704263}}.

\bibitem{Yin:2024:galaxy}
W.~W. Yin, L.~Dai, J.~Huang, L.~Ji, and S.~Ferraro, {\it ``A New Probe of Cosmic Birefringence Using Galaxy Polarization and Shapes''},  2024.

\bibitem{Lue:1999:biref-EB}
A.~Lue, L.~Wang, and M.~Kamionkowski {\em \prl} {\bf 83} (1999) 1506--1509, \href{http://arxiv.org/abs/astro-ph/9812088}{{\tt astro-ph/9812088}}.

\bibitem{Caldwell:2011}
R.~R. Caldwell, V.~Gluscevic, and M.~Kamionkowski {\em \prd} {\bf 84} (2011) 043504, \href{http://arxiv.org/abs/1104.1634}{{\tt arXiv:1104.1634}}.

\bibitem{Lee:2015}
S.~Lee, G.-C. Liu, and K.-W. Ng {\em \plb} {\bf 746} (2015) 406--409, \href{http://arxiv.org/abs/1403.5585}{{\tt arXiv:1403.5585}}.

\bibitem{Leon:2017}
D.~Leon, J.~Kaufman, B.~Keating, and M.~Mewes {\em Mod. Phys. Lett. A} {\bf 32} (2017) 1730002, \href{http://arxiv.org/abs/1611.00418}{{\tt arXiv:1611.00418}}.

\bibitem{Ferreira:2023}
R.~Z. Ferreira, S.~Gasparotto, T.~Hiramatsu, I.~Obata, and O.~Pujolas \href{http://arxiv.org/abs/2312.14104}{{\tt arXiv:2312.14104}}.

\bibitem{Gluscevic:2012}
V.~{Gluscevic}, D.~{Hanson}, M.~{Kamionkowski}, and C.~M. {Hirata} {\em \prd} {\bf 86} (2012) 103529, \href{http://arxiv.org/abs/1206.5546}{{\tt arXiv:1206.5546}}.

\bibitem{PB15:rot}
{\textsc{POLARBEAR} Collaboration} {\em \prd} {\bf 92} (2015) 123509, \href{http://arxiv.org/abs/1509.02461}{{\tt arXiv:1509.02461}}.

\bibitem{BICEP2:2017lpa}
{BICEP2 and Keck Array Collaborations} {\em \prd} {\bf 96} (2017), no.~10 102003, \href{http://arxiv.org/abs/1705.02523}{{\tt arXiv:1705.02523}}.

\bibitem{Contreras:2017}
D.~Contreras, P.~Boubel, and D.~Scott {\em \jcap} {\bf 1712} (2017), no.~12 046, \href{http://arxiv.org/abs/1705.06387}{{\tt arXiv:1705.06387}}.

\bibitem{Namikawa:2020:biref}
T.~Namikawa {\em et~al.} {\em \prd} {\bf 101} (2020), no.~8 083527, \href{http://arxiv.org/abs/2001.10465}{{\tt arXiv:2001.10465}}.

\bibitem{SPT:2020:biref}
F.~Bianchini {\em et~al.} {\em \prd} {\bf 102} (2020), no.~8 083504, \href{http://arxiv.org/abs/2006.08061}{{\tt arXiv:2006.08061}}.

\bibitem{Gruppuso:2020}
A.~Gruppuso, D.~Molinari, P.~Natoli, and L.~Pagano {\em \jcap} {\bf 11} (2020) 066, \href{http://arxiv.org/abs/2008.10334}{{\tt arXiv:2008.10334}}.

\bibitem{Bortolami:2022whx}
M.~Bortolami, M.~Billi, A.~Gruppuso, P.~Natoli, and L.~Pagano {\em \jcap} {\bf 09} (2022) 075, \href{http://arxiv.org/abs/2206.01635}{{\tt arXiv:2206.01635}}.

\bibitem{BK-LoS:2023}
{BICEP2 and Keck Array Collaborations} {\em \apj} {\bf 949} (2023), no.~2 43, \href{http://arxiv.org/abs/2210.08038}{{\tt arXiv:2210.08038}}.

\bibitem{Zagatti:2024}
G.~Zagatti, M.~Bortolami, A.~Gruppuso, P.~Natoli, L.~Pagano, and G.~Fabbian \href{http://arxiv.org/abs/2401.11973}{{\tt arXiv:2401.11973}}.

\bibitem{Li:2014}
S.-Y. Li, J.-Q. Xia, M.~Li, H.~Li, and X.~Zhang {\em \apj} {\bf 799} (2015) 211, \href{http://arxiv.org/abs/1405.5637}{{\tt arXiv:1405.5637}}.

\bibitem{Alighieri:2014yoa}
S.~di~Serego~Alighieri, W.-T. Ni, and W.-P. Pan {\em \apj} {\bf 792} (2014) 35, \href{http://arxiv.org/abs/1404.1701}{{\tt arXiv:1404.1701}}.

\bibitem{Namikawa:2024:BB}
T.~Namikawa {\em \prd} {\bf 109} (4, 2024) 123521, \href{http://arxiv.org/abs/2404.13771}{{\tt arXiv:2404.13771}}.

\bibitem{Kamionkowski:2009:derot}
M.~Kamionkowski {\em \prl} {\bf 102} (2009) 111302, \href{http://arxiv.org/abs/0810.1286}{{\tt arXiv:0810.1286}}.

\bibitem{Namikawa:2016:biref-est}
T.~Namikawa {\em \prd} {\bf 95} (2017), no.~4 043523, \href{http://arxiv.org/abs/1612.07855}{{\tt arXiv:1612.07855}}.

\bibitem{Greco:2022:cross}
A.~Greco, N.~Bartolo, and A.~Gruppuso {\em \jcap} {\bf 03} (2022), no.~03 050, \href{http://arxiv.org/abs/2202.04584}{{\tt arXiv:2202.04584}}.

\bibitem{Greco:2022:aniso-biref-tomography}
A.~Greco, N.~Bartolo, and A.~Gruppuso {\em \jcap} {\bf 05} (2023) 026, \href{http://arxiv.org/abs/2211.06380}{{\tt arXiv:2211.06380}}.

\bibitem{Kamionkowski:1996:eb}
M.~Kamionkowski, A.~Kosowsky, and A.~Stebbins {\em \prd} {\bf 55} (1997) 7368--7388, \href{http://arxiv.org/abs/astro-ph/9611125}{{\tt astro-ph/9611125}}.

\bibitem{Zaldarriaga:1996xe}
M.~Zaldarriaga and U.~Seljak {\em \prd} {\bf 55} (1997) 1830--1840, \href{http://arxiv.org/abs/astro-ph/9609170}{{\tt astro-ph/9609170}}.

\bibitem{Lonappan:LiteBIRD-lens}
A.~I. Lonappan {\em et~al.} {\em \jcap} {\bf 06} (2024) 009, \href{http://arxiv.org/abs/2312.05184}{{\tt arXiv:2312.05184}}.

\bibitem{Namikawa:2012:bhe}
T.~Namikawa, D.~Hanson, and R.~Takahashi {\em \mnras} {\bf 431} (2013) 609--620, \href{http://arxiv.org/abs/1209.0091}{{\tt arXiv:1209.0091}}.

\bibitem{Namikawa:2013:bhe-pol}
T.~Namikawa and R.~Takahashi {\em \mnras} {\bf 438} (2014), no.~2 1507--1517, \href{http://arxiv.org/abs/1310.2372}{{\tt arXiv:1310.2372}}.

\bibitem{Arcari:2024nhw}
S.~Arcari, N.~Bartolo, A.~Greco, A.~Gruppuso, M.~Lattanzi, and P.~Natoli \href{http://arxiv.org/abs/2407.02144}{{\tt arXiv:2407.02144}}.

\bibitem{Hamimeche:2008ai}
S.~Hamimeche and A.~Lewis {\em Phys. Rev. D} {\bf 77} (2008) 103013, \href{http://arxiv.org/abs/0801.0554}{{\tt arXiv:0801.0554}}.

\bibitem{Tristram:2023haj}
M.~Tristram {\em et~al.} {\em Astron. Astrophys.} {\bf 682} (2024) A37, \href{http://arxiv.org/abs/2309.10034}{{\tt arXiv:2309.10034}}.

\bibitem{NASAPICO:2019thw}
{\bf NASA PICO}  {\bf Collaboration} , S.~Hanany {\em et~al.} \href{http://arxiv.org/abs/1902.10541}{{\tt arXiv:1902.10541}}.

\bibitem{Kesden:2002ku}
M.~Kesden, A.~Cooray, and M.~Kamionkowski {\em \prl} {\bf 89} (2002) 011304, \href{http://arxiv.org/abs/astro-ph/0202434}{{\tt astro-ph/0202434}}.

\bibitem{Seljak:2003pn}
U.~Seljak and C.~M. Hirata {\em Phys. Rev. D} {\bf 69} (2004) 043005, \href{http://arxiv.org/abs/astro-ph/0310163}{{\tt astro-ph/0310163}}.

\bibitem{Namikawa:2023:LB-delens}
T.~Namikawa {\em et~al.} {\em \jcap} {\bf 06} (2024) 010, \href{http://arxiv.org/abs/2312.05194}{{\tt arXiv:2312.05194}}.

\end{thebibliography}\endgroup

\end{document}